\documentclass[conference,twocolumn]{IEEEtran}
\usepackage{cite}
\usepackage[T1]{fontenc}
\usepackage{amsmath,amssymb,amsfonts,amsthm}
\usepackage{algorithm}
\usepackage{algorithmic}
\usepackage{graphicx}
\usepackage{booktabs,multirow}
\graphicspath{{./visiofigures/}{./SimFigures/}}
\usepackage{xcolor}
\usepackage{tabularx}
\usepackage{cite}
\usepackage{balance}
\usepackage[unicode=true,
 bookmarks=false,bookmarksnumbered=true,bookmarksopen=false,bookmarksopenlevel=1,
 breaklinks=false,pdfborder={0 0 0},pdfborderstyle={},backref=false,colorlinks=false]
 {hyperref}
\hypersetup{pdftitle={Title},
 pdfauthor={Admin},
 pdfpagelayout=OneColumn, pdfnewwindow=true, pdfstartview=XYZ, plainpages=false}
 
\makeatletter

\theoremstyle{remark}

\makeatother
\allowdisplaybreaks
\begin{document}
\title{Model Aided Deep Learning Based MIMO OFDM Receiver With Nonlinear Power Amplifiers}
\begin{NoHyper}
\author{\IEEEauthorblockN{ Liangyuan~Xu\textsuperscript{$\S \ast$}, Feifei~Gao\textsuperscript{$\S \ast$}, Wei~Zhang\textsuperscript{$\dag $}, and~Shaodan~Ma\textsuperscript{$\ddag $}}  
\IEEEauthorblockA{{\textsuperscript{$\S $} Department of Automation, Tsinghua University, Beijing, China} \\
{\textsuperscript{$\ast $} Beijing National Research Center for Information Science and Technology (BNRist)}\\
{\textsuperscript{$\dag $} School of Electrical Engineering and Telecommunications, University of New South Wales, Sydney, Australia } \\
{\textsuperscript{$\ddag $} Department of Electrical and Computer Engineering, University of Macau, Macao, China } \\
e-mail: \protect\href{mailto:xly18@mails.tsinghua.edu.cn}{xly18@mails.tsinghua.edu.cn}; \protect\href{mailto:feifeigao@ieee.org}{feifeigao@ieee.org}; \protect\href{mailto:w.zhang@unsw.edu.au}{w.zhang@unsw.edu.au}; \protect\href{mailto:shaodanma@um.edu.mo}{shaodanma@um.edu.mo}
}
}
\maketitle
\begin{abstract}
Multi-input multi-output orthogonal frequency division multiplexing (MIMO OFDM) is a key technology for mobile communication systems. However, due to the issue of high peak-to-average power ratio (PAPR), the OFDM symbols may suffer from nonlinear distortions of the power amplifier (PA) at the transmitters, which degrades the channel estimation and detection performances of the receivers. To mitigate the clipping distortions at the receivers end, we leverage deep learning (DL) and devise a DL based receiver which is aided by the traditional least square (LS) channel estimation and the zero-forcing (ZF) equalization  models. Moreover, a data driven DL based receiver without explicit channel estimation is proposed and combined with the model aided DL based receiver to  further improve the performance. Simulation results showcase that the proposed model aided DL based receiver has superior performance of bit error rate and has robustness over different levels of clipping distortions.
\end{abstract}


\section{Introduction}
\IEEEPARstart{M}{ulti}-user multi-input multi-output (MU-MIMO), a promising technology for 5G mobile communication systems, has several appealing advantages, e.g., increasing spatial degree of freedom (DoF)  and enhancing cellular coverage \cite{benefits}. Meanwhile, another standardized  technique in 5G New Radio (NR) is  orthogonal frequency division multiplexing (OFDM) \cite{5GNR},  a popular multi-carrier modulation technique. OFDM has attracted a lot of attention for its  simple implementation and robustness against  frequency-selective fading channel. MIMO OFDM has been shown to provide significant improvement in capacity and spectral efficiency, which makes MIMO OFDM more competitive and attractive.

Although MIMO OFDM has many irresistible features, MIMO OFDM also has many drawbacks, e.g., the beam squint effect in wideband scenario, the significantly large power consumption  of analog-to-digital converters (ADCs), the sensitivity to the carrier frequency offset (CFO) and the issue of high peak-to-average power ratio (PAPR), which motivates many researches \cite{PAPR,WBL,XLY,onebitOFDM,XLYTWC}. Due to the notable issue of high PAPR, the large amplitude parts of the OFDM symbols at the transmitters will force the power amplifier (PA) to work in the nonlinear amplification region, which introduces clipping distortions to the outputs of the PA. The nonlinear distortions then degrade the channel estimation and detection performances at the receivers end. In order to improve the performance of MIMO ODFM receiver, the nonlinear distortions introduced by the PA need to be mitigated.

Recently, the applications of deep learning (DL)  in the  wireless communications physical layer have drawn attentions and earned valuable achievements  over various topics, e.g.,  MIMO channel estimation \cite{HHT,YYWaccess} and MIMO detection \cite{deepMIMO,SISOLY}. DL is also adopted to mitigated the distortions caused by the nonlinear PA. The authors of \cite{DPD} have developed a DL based digital predistortion (DPD) technique for the transmitters to linearize the outputs of the nonlinear PA. However, this is  impractical for mobile devices since they have limited battery and computational resources. In \cite{SISOLY}, the authors have designed a data driven deep neural network (DNN) based single-input single-output (SISO) OFDM receiver which can mitigate clipping distortions. Nevertheless, the performance of this data driven receiver degrades significantly in MIMO OFDM scenario.  In \cite{Echo}, an echo state network based MIMO OFDM receiver has been devised to overcome the nonlinear distortion from the PA. But this sparsely connected network based receiver has poor detection performance and weak robustness. 

In this paper, we propose deep neural network based MIMO OFDM receivers to address the issue introduced by the nonlinear PA. Specifically, the received symbols are first preprocessed by the least square (LS) channel estimation as well as the zero-forcing (ZF) equalization and then fed to a DNN, which is referred to as model aided DL based receiver type {\MakeUppercase{\romannumeral 1}}. Additionally, we propose a data driven receiver where the received symbols are fed to a DNN to accomplish the detection task without explicit channel estimation. Moreover, we propose the model aided DL based receiver type {\MakeUppercase{\romannumeral 2}} which combines the advantages of both the model aided receiver type {\MakeUppercase{\romannumeral 1}} and the data driven receiver. The simulation results show the superior performance of the proposed model aided DL based receiver in terms of bit error  rate.
 
\section{System Model}
\begin{figure}[!tpb]
\centering
\includegraphics[width=0.48\textwidth]{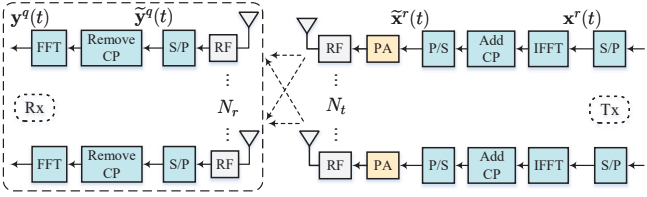}
\caption{Block diagram of MIMO OFDM system.}
\label{fig:syst}
\end{figure}
Consider a MIMO OFDM system  consisting of $N_{r}$ transmit antennas and a base station (BS) with $N_{r}$ receive antennas, as demonstrated in Fig. \ref{fig:syst}. Assume that the number of OFDM subcarriers is $M$, and the maximum length of the channel impulse response is $L$. To combat inter-symbol interference (ISI) caused by the multipath fading channel, a cyclic prefix with length $L_{\rm cp}\geq L-1$ is used. After removing the cyclic prefix, the time-domain signal vector $\widetilde{\mathbf{y}}^{q}(t) \in \mathbb{C}^{M} $ of the $q$th receive antenna in time slot $t$ is given by 
\begin{equation}
\widetilde{\mathbf{y}}^{q}(t)=\sum_{r=1}^{N_{t}} \mathbf{H}_{\mathrm{cir}}^{q, r} \widetilde{\mathbf{x}}^{r}(t) + \widetilde{\mathbf{n}}^{q}(t),
\end{equation}
where $\mathbf{H}_{\mathrm{cir}}^{q, r} \in \mathbb{C}^{M\times M} $ is a circulant matrix with the first column given by $\big[\mathbf{h}^{q, r^{T}}, \mathbf{0}_{1 \times(M - L)}\big]^{T} $, $\mathbf{h}^{q, r} \in \mathbb{C}^{L} $ is the channel impulse response between the $r$th transmit antenna and the $q$th receive antenna, $  \widetilde{\mathbf{x}}^{r}(t) \in \mathbb{C}^{M} $ is the time-domain OFDM symbol transmitted by the $r$th antenna, $\widetilde{\mathbf{n}}^{q}(t) \sim \mathcal{CN}(\mathbf{0},\sigma^2\mathbf{I}_{M})$ is the additive Gaussian noise, and here ${\mathbf I}_{M}$ is an $M\times M$ identity matrix. Then we obtain the frequency-domain symbol of the $q$th receive antenna denoted by $ \mathbf{y}^{q}(t)$ by taking the FFT of $\widetilde{\mathbf{y}}^{q}(t)$, which yields
\begin{equation}
{\mathbf{y}}^{q}(t)= \mathcal{F} \widetilde{\mathbf{y}}^{q}(t)= \sum_{r=1}^{N_{t}} \mathcal{F} \mathbf{H}_{\mathrm{cir}}^{q, r} \widetilde{\mathbf{x}}^{r}(t) + \mathcal{F}  \widetilde{\mathbf{n}}^{q}(t),
\end{equation}
where $\mathcal{F}$ is the $M\times M$ normalized DFT matrix. The eigenvalue decomposition of the circulant matrix $\mathbf{H}_{\mathrm{cir}}^{q, r}$ shows that $ \mathbf{H}_{\mathrm{cir}}^{q, r} = \mathcal{F}^{H} \boldsymbol{\Lambda} \mathcal{F}$, where $\boldsymbol{\Lambda}$ is a diagonal matrix with $\boldsymbol{\Lambda} = \operatorname{diag}\left\{\sqrt{M} \mathcal{F}[\mathbf{h}^{q, r^{T}}, \mathbf{0}_{1 \times M-L}]^{T}\right\}$. Then we have 
\begin{IEEEeqnarray}{ll}\label{eq:OFDM}
{\mathbf{y}}^{q}(t) & = \sum_{r=1}^{N_{t}} \operatorname{diag}\left\{\sqrt{M} \mathcal{F}[\mathbf{h}^{q, r^{T}}, \mathbf{0}_{1 \times M-L}]^{T}\right\} \mathcal{F} \widetilde{\mathbf{x}}^{r}(t) + {\mathbf{n}}^{q}(t)  \notag \\
& = \sum_{r=1}^{N_{t}} \operatorname{diag}\left\{ \mathbf{F}\mathbf{h}^{q, r} \right\} \mathcal{F} \widetilde{\mathbf{x}}^{r}(t) + {\mathbf{n}}^{q}(t),
\end{IEEEeqnarray}
where $ {\mathbf{F}} = \sqrt{M} [\mathcal{F}]_{:,\mathcal{I}(L)}  $, $ {\mathbf{n}}^{q}(t)  = \mathcal{F}\widetilde{\mathbf{n}}^{q}(t)  $, here the notation $[\mathbf{A}]_{:,\mathcal{D}}$ ($[\mathbf{A}]_{\mathcal{D},:}$) denotes the sub-matrix of $\mathbf{A}$ by collecting the columns (rows) indexed by the set $\mathcal{D}$, and $\mathcal{I}(L)$ denotes the set $\{1,2,\cdots,L\}$.

The nonlinearities of PA are ignored in \eqref{eq:OFDM}. The memoryless Rapp model \cite{Rapp,PAsurvey} is widely used to capture the impacts of solid-state nonlinear PA. Assume that $x$ and $x_{\rm PA}$ denotes the input and output signal of PA. The Rapp model can be expressed as
\begin{equation}\label{eq:nonlinPA}
x_{\rm PA} = g(x) = x \frac{\Gamma(|x|)\exp(\jmath \phi(|x|))}{|x|},
\end{equation}
where $\Gamma(|x|)$ is the AM/AM conversion (the amplitude transfer characteristics), $\phi(|x|)$ is the AM/PM conversion (the phase transfer characteristics), and $\jmath = \sqrt{-1}$. $\Gamma(|x|)$ and $\phi(|x|)$ are given by \cite{PAsurvey}
\begin{equation}
\Gamma(|x|)=|x|\left(1+\left(\frac{|x|}{\nu_{\rm sat }}\right)^{2 \delta}\right)^{-\frac{1}{2 \delta}}, \quad \phi(|x|) = 0,
\end{equation}
where $\nu_{\rm sat }$ is the amplitude of the saturated output (clipping threshold), and $\delta$ is the smooth factor (we set $\delta =5$). The severity of the distortion depends on the \emph{clipping level} defined as $\nu^{2}_{\rm sat}/\rho$, where $\rho$ is the average input power of the PA. 

With the existence of distortions caused by nonlinear PA, we adopt the Rapp model to rewrite the received symbol in \eqref{eq:OFDM} as 
\begin{IEEEeqnarray}{ll}\label{eq:PAOFDM}
{\mathbf{y}}^{q}(t) & = \sum_{r=1}^{N_{t}} \operatorname{diag}\left\{ \mathbf{F}\mathbf{h}^{q, r} \right\} \mathcal{F} g\Big(\widetilde{\mathbf{x}}^{r}(t)\Big) + {\mathbf{n}}^{q}(t) \notag \\
& = \sum_{r=1}^{N_{t}} \operatorname{diag}\left\{ \mathbf{F}\mathbf{h}^{q, r} \right\} \mathcal{F} g\Big( \mathcal{F}^{H} \mathbf{x}^{r}(t)\Big) + {\mathbf{n}}^{q}(t),
\end{IEEEeqnarray}
where $g(\cdot)$ given by \eqref{eq:nonlinPA} is the element-wise characteristics function of nonlinear PA, and $\mathbf{x}^{r}(t)$ denotes the frequency-domain OFDM symbol transmitted by the $r$th antenna.
\section{Model Aided Deep Learning Based Receiver}
\begin{figure}[!tpb]
\centering
\includegraphics[width=0.47\textwidth]{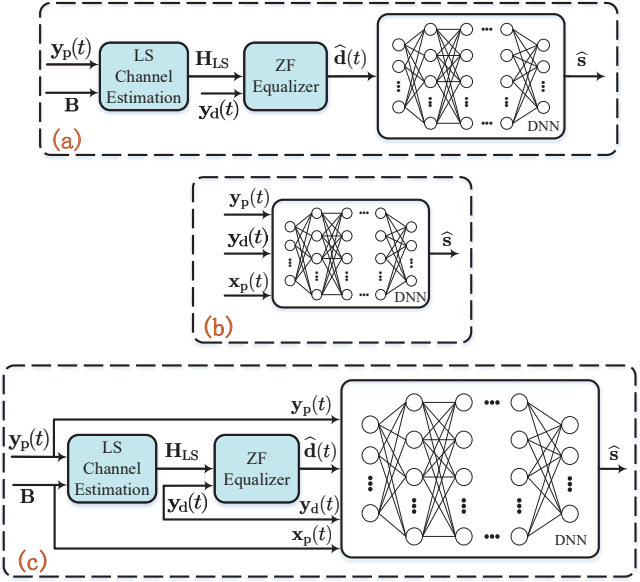}
\caption{The proposed MIMO OFDM receivers: (a) model aided DL based receivers type {\MakeUppercase{\romannumeral 1}}; (b) data driven receiver; (c) model aided DL based receivers type {\MakeUppercase{\romannumeral 2}}.}
\label{fig:receivers}
\end{figure}
\subsection{Model Aided Deep Learning Based Receiver Type {\MakeUppercase{\romannumeral 1}}}\label{sec:type1}
We use subscript-$\rm p$ and $\rm d$ to denote the pilots transmission and the payload data transmission, respectively.  During pilot symbols transmission, we rewrite \eqref{eq:PAOFDM} as
\begin{IEEEeqnarray}{ll}
{\mathbf{y}}^{q}_{\rm p}(t) & = \sum_{r=1}^{N_{t}} \operatorname{diag}\left\{ \mathbf{F}_{\rm p} \mathbf{h}^{q, r} \right\} \mathcal{F}_{\rm p} g\Big(\mathcal{F}^{H} \mathbf{x}^{r}(t)\Big)  + {\mathbf{n}}^{q}_{\rm p} (t) \notag \\
& = \mathbf{A} \mathbf{h}^{q}  + {\mathbf{n}}^{q}_{\rm p} (t),
\end{IEEEeqnarray}
where we assume the number of pilots in one OFDM symbol is $M_{\rm p}$, the subcarriers of pilots are indexed by the set $ \mathcal{D}_{\rm p} $, $ \mathbf{y}_{\rm p}^{q}(t) \in \mathbb{C}^{M_{\rm p}} $ is the received vector of the corresponding subcarriers, $ {\mathbf{F}}_{\rm p} = [\mathbf{F}]_{ \mathcal{D}_{\rm p} ,:}  $, $ {\mathcal{F}}_{\rm p} = [\mathcal{F}]_{ \mathcal{D}_{\rm p} ,:}  $, $\mathbf{h}^{q} = \big[\mathbf{h}^{q, 1^{T}}, \ldots, \mathbf{h}^{q, N_{t}^{T}}\big]^{T} $, and $\mathbf{A}$ is given by
\begin{multline}
 \mathbf{A} = \big[ \operatorname{diag}\left\{ \mathcal{F}_{\rm p} g\big(\mathcal{F}^{H} \mathbf{x}^{1}(t)\big) \right\} \mathbf{F}_{\rm p} ,\cdots,\\
 \operatorname{diag}\left\{ \mathcal{F}_{\rm p} g\big(\mathcal{F}^{H} \mathbf{x}^{r}(t)\big) \right\} \mathbf{F}_{\rm p} \big].
\end{multline}
If $\mathbf{A}$ is perfectly known by the receiver, the LS  channel estimation can be obtain by $\mathbf{A}^{\dag} {\mathbf{y}}^{q}_{\rm p}(t) $, where the notation $(\cdot)^\dagger$ denotes matrix  pseudo inverse. However, the receiver has no knowledge of PA nonlinearities, and hence the receiver cannot obtain $\mathbf{A}$ perfectly and only knows
\begin{equation}
 \mathbf{B} = \left[ \operatorname{diag}\left\{ \mathbf{x}^{1}_{\rm p} (t) \right\} \mathbf{F}_{\rm p} ,\cdots,\operatorname{diag}\left\{ \mathbf{x}^{r}_{\rm p} (t) \right\} \mathbf{F}_{\rm p} \right],
\end{equation}
where $ \mathbf{x}^{r}_{\rm p} (t) \in \mathbb{C}^{M_{\rm p}} $ is the pilots. Then, the LS  channel estimation is given by 
\begin{equation}
\widehat{\mathbf{h}}^{q}_{\rm LS} = \mathbf{B}^{\dag} {\mathbf{y}}^{q}_{\rm p}(t) = \mathbf{B}^{\dag} \mathbf{A} \mathbf{h}^{q} + \mathbf{B}^{\dag} {\mathbf{n}}^{q}(t).
\end{equation}

During payload data transmission, according to \eqref{eq:PAOFDM}, the received OFDM symbol ${\mathbf{y}}^{q}_{\rm d} (t) \in \mathbb{C}^{M} $ can be expressed as 
\begin{IEEEeqnarray}{ll}\label{eq:yq}
{\mathbf{y}}^{q}_{\rm d} (t)  & = \sum_{r=1}^{N_{t}} \operatorname{diag}\left\{ \mathbf{F}\mathbf{h}^{q, r} \right\} \mathcal{F} g\Big( \mathcal{F}^{H} \mathbf{x}_{\rm d}^{r}(t) \Big) + {\mathbf{n}}^{q}_{\rm d} (t) \notag \\
& =  \mathbf{H}^{q} \mathbf{d}(t) + {\mathbf{n}}^{q}_{\rm d} (t),  
\end{IEEEeqnarray}
where $\mathbf{x}_{\rm d}^{r}(t) \in \mathbb{C}^{M} $ is the payload data symbol, $\mathbf{H}^{q} = \left[ \operatorname{diag}\left\{ \mathbf{F}\mathbf{h}^{q, 1} \right\} \mathcal{F} ,\cdots, \operatorname{diag}\left\{ \mathbf{F}\mathbf{h}^{q, N_{t}} \right\} \mathcal{F} \right]$, and $\mathbf{d}(t) = \big[ g\big( \mathcal{F}^{H} \mathbf{x}_{\rm d}^{1}(t) \big)^{T} ,\ldots, g\big(\mathcal{F}^{H} \mathbf{x}_{\rm d}^{N_{t}}(t) \big)^{T} \big]^{T} $. Combining $N_{r}$ symbol vectors of \eqref{eq:yq} into a long vector yields 
\begin{equation}
{\mathbf{y}}_{\rm d} (t) = \mathbf{H} \mathbf{d}(t) + {\mathbf{n}}_{\rm d} (t), 
\end{equation}
where ${\mathbf{y}}_{\rm d} (t)=[{\mathbf{y}}^{1}_{\rm d} (t)^{T},\cdots,{\mathbf{y}}^{N_{r}}_{\rm d} (t)^{T}]^{T}$, $\mathbf{H}=\big[{\mathbf{H}^{q}}^{T},\cdots,{\mathbf{H}^{N_{r}}}^{T}\big]^{T}$ and ${\mathbf{n}}_{\rm d} (t) = [{\mathbf{n}}^{1}_{\rm d} (t)^{T},\cdots,{\mathbf{n}}^{N_{r}}_{\rm d} (t)^{T}]^{T}$. Substituting $\widehat{\mathbf{h}}^{q}_{\rm LS}$ into $\mathbf{H}$, we have $ {\mathbf H }_{\rm LS} $ and then obtain the zero-forcing equalization of $\mathbf{d}(t)$ as follows:
\begin{equation}
\widehat{\mathbf{d}}(t) = \mathbf{H}^{\dag}_{\rm LS} {\mathbf{y}}_{\rm d}(t) = \mathbf{C} \mathbf{d}(t) + \mathbf{e}_{\rm d}(t),
\end{equation}
where $\mathbf{C} \triangleq \mathbf{H}^{\dag}_{\rm LS} \mathbf{H} $ and $\mathbf{e}_{\rm d}(t) \triangleq \mathbf{H}^{\dag}_{\rm LS} { \mathbf{n}}_{\rm d}(t)$. 

After obtaining $\widehat{\mathbf{d}}(t)$, we should find a nonlinear mapping $\Phi:\{\widehat{\mathbf{d}}(t)\}\rightarrow\{\mathbf{x}_{\rm d}(t)\}$ to eliminate the impacts of $\mathbf{C}$ as well as $g(\cdot)$ and accomplish the symbol detection task where $\mathbf{x}_{\rm d}(t) = \big[\mathbf{x}_{\rm d}^{1}(t)^{T},\cdots,\mathbf{x}_{\rm d}^{N_{t}}(t)^{T}\big]^{T}$. However, the receiver cannot find the mapping function easily for the lack of knowledge about the matrix $\mathbf{C}$ and PA characteristics, which motivates us to resort to DL method to find the nonlinear mapping. Since we first obtain $\widehat{\mathbf{d}}(t)$ aided by LS channel estimation as well as ZF equalization model and then leverage DL to accomplish the detection task, we name this receiver as model aided DL based receiver type {\MakeUppercase{\romannumeral 1}}. The block diagram of this receiver is demonstrated in Fig. \ref{fig:receivers}(a). As can be seen,  $\widehat{\mathbf{d}}(t)$ (the real part and imaginary part) is fed to a DNN with $\mathcal{L}$ layers (one input layer, $\mathcal{L}-1$ hidden layers and one output layer). The outputs of the DNN $\widehat{\mathbf{s}}$ is a cascade of nonlinear transformation of the input, which yields
\begin{equation}
\widehat{\mathbf{s}} =f^{(\mathcal{L})}\left(\mathbf{W}^{(\mathcal{L})}f^{(\mathcal{L}-1)}\big(\cdots f^{(2)}(\mathbf{W}^{(2)}\widehat{\mathbf{d}} + \mathbf{b}^{(2)})\big) + \mathbf{b}^{(\mathcal{L})}\right), \notag
\end{equation} 
where $f^{(\ell)}$, $\mathbf{W}^{(\ell)}$ and $\mathbf{b}^{(\ell)}$ are the activation function, weight matrix and bias vector of the $\ell$th layer  $\forall\ell=2,\cdots,\mathcal{L}$, respectively. We define $\boldsymbol{\theta} \triangleq \{\mathbf{W}^{(\ell)},\mathbf{b}^{(\ell)}\}^{\mathcal{L}}_{\ell=2}$, which are the trainable parameters of the network. The ReLU function is used as the activation function of hidden layers ($\ell=2,\cdots,\mathcal{L}-1$) of the DNN. The output layer ($\ell=\mathcal{L}$) of the DNN takes the Sigmoid function as the activation function to map the output into the interval $[0,1]$.

The outputs of each DNN are the detected bits data of $K$ carriers, which means there are total $N_{t}M/K$ DNNs. These DNNs  are trained independently, and the outputs of these DNNs are concatenated for the detected bit streams of $\mathbf{x}_{\rm d}(t)$.  Moreover, the loss function of each DNN is 
\begin{equation}
\operatorname{LOSS}(\boldsymbol{\theta}) = \frac{1}{VP}\sum_{v=1}^{V}\Vert\widehat{\mathbf{s}}^{(v)}-\mathbf{s}^{(v)}\Vert^{2}_{2},
\end{equation}
where $V$ is the batch size \footnote{Batch size is the number of samples in one training batch.}, $P$ is the length of the vector $\mathbf{s}$, $\widehat{\mathbf{s}}$ is the output of the DNN, $\mathbf{s}$ is the supervision label, and the superscript-$(v)$ is the index in the training batch.

\subsection{Data Driven Deep Learning Based Receiver}\label{sec:datadriven}
In the above subsection, some preprocessing of the received symbol is aided by LS channel estimation and ZF equalization. Then DL is leveraged to mitigate the residual distortions and accomplish the detection task. Instead of explicitly estimating and equalizing the wireless channel, we can apply DL in end-to-end manner. We are supposed to find the following mapping without explicit channel estimation
\begin{equation}
\Psi:\{\mathbf{y}_{\rm p}(t),\mathbf{x}_{\rm p} (t),{\mathbf{y}}_{\rm d} (t)\}\rightarrow\{\mathbf{x}_{\rm d}(t)\},
\end{equation}
where $\mathbf{y}_{\rm p}(t) = \big[\mathbf{y}^{1}_{\rm p}(t)^{T},\cdots,\mathbf{y}^{N_{r}}_{\rm p}(t)^{T}\big]^{T}$ is the received pilot symbol, $\mathbf{x}_{\rm p} (t) = \big[\mathbf{x}^{1}_{\rm p} (t)^{T},\cdots,\mathbf{x}^{N_{t}}_{\rm p} (t)^{T}\big]^{T}$ is the pilot symbol.  The universal approximation theorem \cite{UAT} shows the powerful representational capacities of DL. Hence, with  supervised learning and sufficient training, DL is able to find such a mapping $\Psi$. We call this as data driven DL based receiver whose block diagram is illustrated in Fig. \ref{fig:receivers}(b). The vector $\big[\mathbf{y}_{\rm p}(t)^{T},\mathbf{x}_{\rm p} (t)^{T},{\mathbf{y}}_{\rm d} (t)^{T}\big]^{T}$ (the real part and imaginary part) is fed to a DNN with $\mathcal{L}$ layers. The outputs of the DNN are the detected bits data of $K$ carriers as the same with Section \ref{sec:type1}, which means there are $N_{t}M/K$ independently trained DNNs. The loss function is also the same with Section \ref{sec:type1}.

Note that the pilot symbols are necessary for the data driven receiver to avoid detection ambiguity, since there is no explicit channel equalization.

\subsection{Model Aided Deep Learning Based Receiver Type {\MakeUppercase{\romannumeral 2}}}
The main difference between the receiver in  \ref{sec:type1} and the receiver in \ref{sec:datadriven} is whether there is some model aided preprocessing of the DNN input. Compare to the data driven receiver in \ref{sec:datadriven}, the main advantages of the model aided DL receiver in \ref{sec:type1} are given as follows: (\romannumeral 1) the dimension of the DNN input can be reduced ($\widehat{\mathbf{d}}(t)\in \mathbb{C}^{N_{t}M} $, ${\mathbf{y}}_{\rm d} (t) \in \mathbb{C}^{N_{r}M}$ and $N_{r}>N_{t}$ for most MIMO scenario); (\romannumeral 2) the residual distortions in $\widehat{\mathbf{d}}(t)$ is small in the high SNR regime, which means the DNN is easy to be trained and would have better performance. However, in the low SNR regime, the performance of the model aided receiver is likely to be poor due to the error propagation. Here the error propagation means that the LS channel estimation error will propagate to $\widehat{\mathbf{d}}(t)$ through ZF equalization, which dominates the residual distortions in $\widehat{\mathbf{d}}(t)$ for the low SNR regime.  The proposed data driven receiver will not suffer from error propagation, since it has no explicit channel estimation and equalization.

To take full advantages of both the receiver in  \ref{sec:type1} and in \ref{sec:datadriven}, we combine the $\widehat{\mathbf{d}}(t)$ and $\big[\mathbf{y}_{\rm p}(t)^{T},\mathbf{x}_{\rm p} (t)^{T},{\mathbf{y}}_{\rm d} (t)^{T}\big]^{T}$ as the input of a DNN with $\mathcal{L}$ layers. The outputs of the DNN are the detected bits data of $K$ carriers as the same with Section \ref{sec:type1} and \ref{sec:datadriven}. The loss function is also the same with Section \ref{sec:type1} and \ref{sec:datadriven}. We name this receiver as model aided DL based receiver type {\MakeUppercase{\romannumeral 2}} whose block diagram is illustrated in Fig. \ref{fig:receivers}(c).
\section{Numerical Results}
\begin{table*}[!tbh]
\centering
\caption{Network configurations  of the proposed DL based receivers}
\label{table:network}
\begin{tabular}{cccc}
\toprule
 & Model Aided Receivers Type {\MakeUppercase{\romannumeral 1}} & Data Driven Receivers & Model Aided Receivers Type {\MakeUppercase{\romannumeral 2}} \\
\midrule
Input Layer & $2N_{t}M$  & $2(N_{r}M_{\rm p}+N_{t}M_{\rm p}+N_{r}M)$  & $2(N_{r}M_{\rm p}+N_{t}M_{\rm p}+N_{r}M+N_{t}M)$  \\
\cmidrule{1-4}
Dense Layer 1 & 1024, Batch Normalization, ReLU & 4000, Batch Normalization, ReLU & 4000, Batch Normalization, ReLU\\
\cmidrule{1-4}
Dense Layer 2  & 2028, Batch Normalization, ReLU & 3000, Batch Normalization, ReLU & 3000, Batch Normalization, ReLU\\
\cmidrule{1-4}
Dense Layer 3  & 512, Batch Normalization, ReLU & 1024, Batch Normalization, ReLU & 1024, Batch Normalization, ReLU\\
\cmidrule{1-4}
Output Layer  & 32, Batch Normalization, sigmoid & 32, Batch Normalization, sigmoid & 32, Batch Normalization, sigmoid\\
\bottomrule
\end{tabular}
\end{table*}

We conduct several simulations to showcase the bit error rate performances of the proposed model aided DL based receivers and the data driven receiver. The $N_{t}$ transmit antennas will send different data sequences. The pilot sequence is chosen as a length-$M_{p}=N_{t}L$ sequence, and the pilot for each transmit antenna is a constant amplitude orthogonal sequence. The pilot tones are equispaced within the OFDM symbol. The networks output the detected bits data of $K=8$ carriers, and 16QAM modulation scheme is used, which indicates the size of output layer is 32. The channel is normalized, and SNR is
\begin{equation}
\mathrm{SNR} \triangleq 10\log_{10}\left( \frac{ N_{t}\rho }{ \sigma^2} \right)\quad \mathrm{dB},
\end{equation}
where $\rho $ is the average transmission power of the users. The number of carrier is $M=128$, and the maximum length of the channel response is $L=16$. The channel impulse response is generated by the MATLAB toolbox of MIMO rayleigh fading channel model. The number of channel paths is randomly chosen within $10$, and maximum delay $6$ sampling period is chosen. We assume that the channel remains constant during the pilots  and payload data transmission. The network configurations of the proposed receivers are listed in Table \ref{table:network}, where these networks are optimized by ADAM \cite{adam} with initial learning rate $0.001$. The training set contains 2.4e5 samples, and the batch size is 300.

In the simulations, the performances of the following algorithms are given as benchmarks:
\begin{itemize}
\item The traditional receiver algorithm with  LS channel estimation and ZF detection, where ideal linear PAs are deployed at transmitters. We use "LS+ZF, Linear PA" to denote this algorithm.
\item LS channel estimation and ZF detection algorithm, where nonlinear PAs are deployed at transmitters. However, the receiver has no knowledge of PA nonlinearities and assume PAs are  ideal linear. We use "LS+ZF, Nonlinear PA" to denote this algorithm.
\item The receiver algorithm with  LS channel estimation and maximum-likelihood detection (MLD), where nonlinear PAs are deployed at transmitters. Here, the nonlinear function is  perfectly known by the receiver, and this perfect knowledge of  PA nonlinearities is utilized in both LS estimation and MLD. Note that MLD is applied to all the $M$ carriers jointly, which is computationally prohibitive ($4^M$ searches). To address this issue, we apply MLD to the corresponding $K$ carriers by fixing the data of the remaining $M-K$ carriers. If the data of the remaining $M-K$ carriers is  randomly generated and then fixed, we use "LS+MLD, Upper Bound"  to denote this case. If we assume the data of the remaining $M-K$ carriers is correct, we use "LS+MLD, Lower Bound"  to denote this case.  
\end{itemize}

\begin{figure}[!tpb]
\centering
\includegraphics[width=0.45\textwidth]{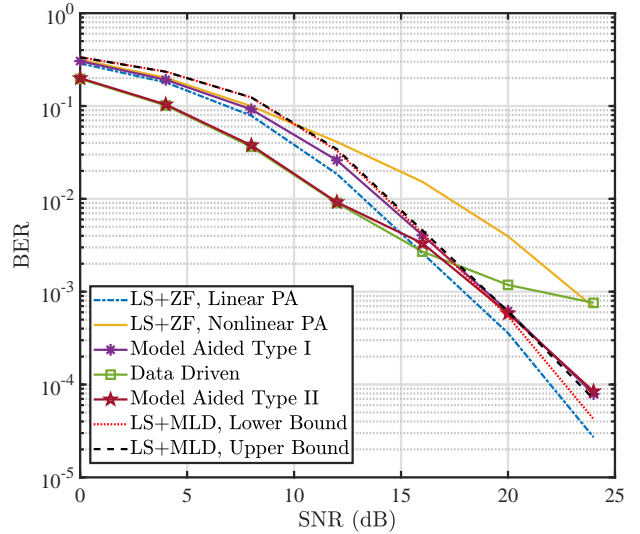}
\caption{BER versus SNR for the proposed receivers, where $N_{t}=2$, $N_{r}=8$ and the clipping level of the nonlinear PAs is 7dB.}
\label{fig:2x8Rapp7dB}
\end{figure}
In Fig. \ref{fig:2x8Rapp7dB}, the BER of the proposed receivers is plotted over SNR, where $N_{t}=2$, $N_{r}=8$ and the clipping level of the nonlinear PAs is 7dB.  As can been seen, the performance of "LS+ZF, Nonlinear PA" is surpassed by "LS+MLD, Lower Bound" and "LS+MLD, Upper Bound". The reason is that the receivers have  leveraged the perfect knowledge of  PA nonlinearities to compensate the receive algorithm in "LS+MLD, Lower Bound" and "LS+MLD, Upper Bound". Additionally, the performance of the proposed model aided receiver type {\MakeUppercase{\romannumeral 1}} is very close to "LS+MLD, Lower Bound" and superior to "LS+ZF, Nonlinear PA", which indicates that the proposed model aided receiver type {\MakeUppercase{\romannumeral 1}} is able to  learn the features of nonlinearities and mitigate nonlinear distortions with the assistance of DL.  Moreover, the proposed data driven receiver has the best performance in the low SNR regime but has poor performance in the high SNR regime. On the one hand, the reason is that the data driven receiver has no explicit channel estimation as well as equalization and hence will not suffer from error propagation which is dominant in the low SNR regime. On the other hand, without the assistance of explicit channel equalizer, the data driven receiver cannot properly overcome inter-user interference which is dominant in the high SNR regime. Furthermore, the proposed model aided receiver type {\MakeUppercase{\romannumeral 2}} offers good performances in both low and high SNR regimes,  since it combines the advantages of both the receiver type {\MakeUppercase{\romannumeral 1}} and the data driven receiver.

\begin{figure}[!tpb]
\centering
\includegraphics[width=0.45\textwidth]{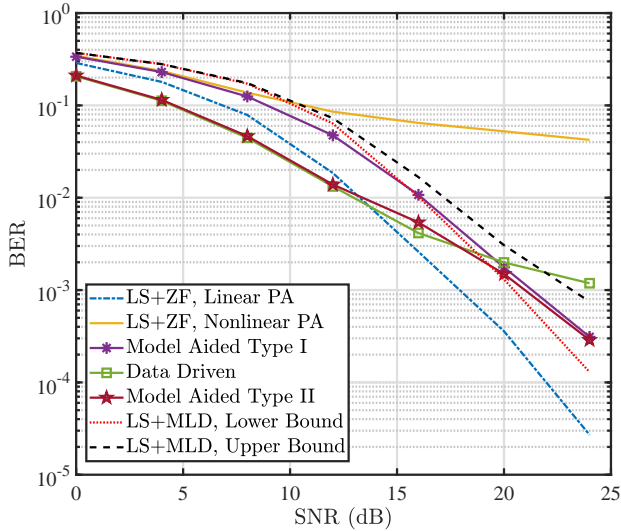}
\caption{BER versus SNR for the proposed receivers, where $N_{t}=2$, $N_{r}=8$ and the clipping level of the nonlinear PAs is 5dB.}
\label{fig:2x8Rapp5dB}
\end{figure}
Fig. \ref{fig:2x8Rapp5dB} displays the BER over SNR, where $N_{t}=2$, $N_{r}=8$.  To showcase the robustness of the proposed receivers, the  nonlinear distortions of the PAs are more severe with 5dB clipping level.  We see that the algorithm of "LS+ZF, Nonlinear PA" fails to work properly due to the severe distortions introduced by the nonlinear PAs. Nevertheless, the performance the proposed model aided receiver type {\MakeUppercase{\romannumeral 1}} is still very close to "LS+MLD, Lower Bound", and  the proposed data driven receiver has the best performance in the low SNR regime, and the proposed model aided receiver type {\MakeUppercase{\romannumeral 2}} still works properly and takes full advantages of both the receiver type {\MakeUppercase{\romannumeral 1}} and the data driven receiver. These results showcase the robustness and effectiveness of the proposed receivers.

\section{Conclusions}
In this paper, DL is adopted to address the issue of nonlinear distortions introduced by the PAs of the transmitters in MIMO OFDM system.  We proposed a model aided DL based receiver type {\MakeUppercase{\romannumeral 1}} assisted by the LS channel estimation as well as ZF equalization,  and a data driven receiver without explicit channel estimation. To further improve the performance, we devise a DL based receiver which combines the advantages of both the receiver type {\MakeUppercase{\romannumeral 1}} and the data driven receiver. Numerical results showcase the robustness and superior performances of the proposed receivers, and support the application of the proposed receivers in MIMO OFDM system. 

\section*{Acknowledgment}
This work was supported in part by the National Natural Science Foundation of China under Grant 61831013, Grant 61771274, and Grant 61531011, by the Beijing
Municipal Natural Science Foundation under Grant 4182030 and Grant L182042, by Shenzhen Science and Innovation Fund under Grant JCYJ20180507182451820 and Grant JCYJ20170412104656685, by the Science and Technology Development Fund of Macau SAR under Grant 0036/2019/A1 and Grant SKL-IOTSC2018-2020, and by the Research Committee of University of Macau under Grant MYRG2018-00156-FST. 

\balance
\bibliographystyle{IEEEtran}
\bibliography{IEEEabrv,PApaper}
\end{NoHyper}
\end{document}